\documentclass[prd, aps, nofootinbib, preprint, letterpaper]{revtex4-2}

\usepackage{xcolor}
\usepackage[export]{adjustbox}
\usepackage{appendix}
\usepackage{extarrows}
\usepackage{amsmath, amssymb, amsfonts}
\usepackage{mathtools}
\usepackage{mathrsfs}

\usepackage[hidelinks]{hyperref}
\usepackage{ulem}
\usepackage{setspace}
\usepackage{subcaption} 
\usepackage{comment}

\newcommand{\checked}[1]{}
\newcommand{\beq}{\begin{equation}}
\newcommand{\eeq}{\end{equation}}
\newcommand{\bqa}{\begin{eqnarray}}
\newcommand{\eqa}{\end{eqnarray}}

\begin{document}

\title {Deconfining Phase Transition under Real Rotation: A Matrix Model Study}

\author{Qianqian Du$^{a,b}$, Jing He$^{a}$, Yun Guo$^{a,b,*}$, Mei Huang$^{c,\dagger}$, Enke Wang$^{d,e,\ddagger}$}

\affiliation{
$^a$ Department of Physics, Guangxi Normal University, Guilin, 541004, China\\\vspace{-3pt}
$^b$ Guangxi Key Laboratory of Nuclear Physics and Technology, Guilin, 541004, China\\\vspace{-3pt}
$^c$School of Nuclear Science and Technology, University of Chinese Academy of Sciences, Beijing, 100049, China\\\vspace{-3pt}
$^d$ State Key Laboratory of Nuclear Physics and Technology, Institute of Quantum Matter,
South China Normal University, Guangzhou 510006, China\\ \vspace{-3pt}
$e$ Guangdong Basic Research Center of Excellence for Structure and Fundamental Interactions of Matter, Guangdong Provincial Key Laboratory of Nuclear Science, Guangzhou 510006, China\vspace{-3pt}
}

\renewcommand{\thefootnote}{\fnsymbol{footnote}}
\footnotetext[1]{ yunguo@mailbox.gxnu.edu.cn}

\footnotetext[2]{
huangmei@ucas.ac.cn}

\footnotetext[3]{
wangek@scnu.edu.cn}
\renewcommand{\thefootnote}{\arabic{footnote}}

\begin{abstract} 

We construct a matrix model to study the deconfining phase transition for a pure gluon plasma that is confined in a cylinder of radius ${\cal R}$ and rotating rigidly at a real-valued angular velocity $\Omega$, satisfying $\mathcal{R} \Omega<1$. The deconfining phase transition arises due to the competition between two terms that constitute the matrix model. The perturbative term comes from the one-loop effective potential computed in the presence of a background field, while the non-perturbative term represents a correction to the perturbative contribution which is brought about by taking into account an effective mass of the gauge fields.
Our results show that real rotation induces a radial inhomogeneity of the system and the deconfining temperature $T_c$ drops away from the rotation axis which is consistent with the Tolman-Ehrenfest law. As for the $\Omega$-dependence of $T_c$, it relies on our assumptions of the gluon effective mass. For a constant mass, $T_c$ is found to always decrease with increasing $\Omega$. A non-monotonic behavior of $T_c$ shows up when a $\Omega$-dependent mass is considered, leading to a qualitative change in the region of small angular velocity. In addition, by setting $\Omega=0$ to eliminate rotational effects, we also demonstrate that the finite-volume effect reduces the deconfining temperature relative to the infinite-volume limit. Comparisons between our results and those from various lattice simulations and phenomenological models suggest that controversy remains over how the deconfining phase transition is modified by real rotation and further work is required to reach a definite conclusion.

\end{abstract}

\maketitle
\newpage 
 
\section{Introduction}

High-energy heavy-ion collisions conducted at the Large Hadron Collider (LHC) and the Relativistic Heavy Ion Collider (RHIC) produce a state of matter called quark-gluon plasma (QGP)~\cite{Gyulassy:2004zy,Shuryak:2004cy}. In these experiments, collisions between two nucleis are rarely head-on, those non-central collisions lead to the rapid rotation of QGP with angular momentum perpendicular to the collision plane~\cite{Becattini:2007sr,Baznat:2013zx,STAR:2017ckg,Jiang:2016woz,Deng:2016gyh,Becattini:2020ngo}.
The value of the angular velocity $\Omega \approx (9\pm 1) \times 10^{21} \, \rm{s}^{-1} \thicksim 6\, \rm{MeV} $ was first reported by the STAR collaboration at RHIC~
\cite{STAR:2017ckg} which indicates that QGP is the fastest-rotating fluid ever observed in nature.
The theoretical analysis predicts even a faster angular velocity depending on the initial parameters of the collision. For example, an estimate for $\Omega$ based on a multiphase transport (AMPT) model can reach $\thicksim 20-40\, \rm{MeV}$~\cite{Jiang:2016woz}. We can thus expect that the QGP produced in heavy-ion collisions have a very strong vortex structure. 

The sufficiently rapid rotation influences the properties of QGP and leads to a series of fascinating physical phenomena that can be observed in experiments. 
Rotation induces spin polarization through spin-orbital coupling~\cite{Huang:2011ru}, 
including not only the global polarization of hyperons~\cite{Liang:2004ph,STAR:2022fan}, but also the spin alignment of vector mesons~\cite{ALICE:2019aid}.
In a system of chiral fermions, rotation plays a role closely analogous to that of an external magnetic field, resulting in the anomalous transport phenomena known as chiral vortical effect~\cite{Son:2009tf,Kharzeev:2010gr,Prokhorov:2018bql} and chiral vortical wave~\cite{Rogachevsky:2010ys,Baznat:2013zx,Jiang:2015cva,Becattini:2013vja}.  
In addition, rotation gives rise to the Kelvin-Helmholtz instability~\cite{Csernai:2011qq} in relativistic heavy-ion collisions, and generates distinctive nuclear states, such as rotational modes and high-spin states~\cite{deVoigt:1983zz,Garrett:1986in} in low-energy nuclear physics. 

It is also of interest to study the effect of rotation on the QCD phase transition. 
On the one hand, results for chiral phase transition from both lattice simulations~\cite{Braguta:2022str} and phenomenological studies~\cite{Wang:2018sur,Ebihara:2016fwa,Ebihara:2016fwa,  Chen:2023cjt,Chernodub:2016kxh} 
converge on the conclusion that rotation suppresses the chiral condensate and lowers the critical temperature.
This is because rotation tends to align the spins of quarks and antiquarks along the rotation axis, and thus inhibits the scalar pairing~\cite{Jiang:2016wvv,Chernodub:2022veq} which can be considered as a generalization of the Barnett effect~\cite{Barnett:1915uqc}. On the other hand, there is no definite conclusion regarding the effect of rotation on deconfining phase transition. After analytically continuing to the real-valued angular velocity, most of the lattice simulations suggest that rotation increases the deconfining temperature~\cite{Braguta:2020biu, Braguta:2021jgn,Braguta:2021ucr,Braguta:2023iyx}, while the lattice result using the strong coupling expansion method predicts an opposite behavior~\cite{Wang:2025mmv}. There are also extensive theoretical and modeling studies on this topic, including holographic QCD~\cite{Wang:2024szr,Chen:2020ath,Braga:2022yfe,Zhao:2022uxc,Chen:2024jet,Chen:2024edy,Li:2026pkd}, hadron resonance gas model~\cite{Fujimoto:2021xix}, an analytically tractable model in compact QED~\cite{Chernodub:2020qah}, the caloron model~\cite{Jiang:2024zsw,Jiang:2023zzu}, and the bag model~\cite{Mameda:2023sst}. Furthermore, the phase transition of a rotating QGP has been also studied using the Polarized-Polyakov-loop Nambu-Jona-Lasinio model which also takes into account the effects of rotating gluons on the quark polarization \cite{Sun:2024anu}. However, due to different theoretical frameworks and model assumptions, no convergent conclusion can be drawn at present, although most of the studies support a decrease in the deconfining temperature with increasing angular velocity.
In addition, both lattice simulations~\cite{Braguta:2023iyx} and model studies~\cite{Chernodub:2020qah,Jiang:2024zsw} show that the rotating plasma becomes inhomogeneous because the deconfining temperature also varies with the radial distance from the axis of rotation, however, whether such an inhomogeneity obeys the Tolman-Ehrenfest law~\cite{Tolman:1930ona} remains an unresolved inconsistency. These discrepancies among existing results  reveal gaps in our current understanding and necessitate further investigation into the deconfining phase transition under rotation. 

The matrix model for deconfinement was first proposed in Ref.~\cite{Meisinger:2001cq} where a non-trivial background field is introduced to describe the temperature dependence of the order parameter, i.e., the Polyakov loop in the deconfining phase transition. It has been further developed~\cite{Dumitru:2012fw,Dumitru:2010mj,Guo:2014zra} to quantitatively reproduce the lattice data of the thermodynamics in the so-called semi-QGP region~\cite{Hidaka:2008dr}. The matrix model was also extended to investigate the chiral phase transition by including the fermionic contributions~\cite{Pisarski:2016ixt} and generalized to discuss other related topics~\cite{Ren:2026xfn,Du:2024riq,Liu:2024fki,Guo:2020jvc,Lin:2013efa,Gale:2014dfa,Debnath:2025qhd}. Given the success of the matrix model in studying the QCD phase transition, in this work, we will make use of it to explore the deconfining
phase transition by taking into account the rotation effect for both $SU(2)$ and $SU(3)$ gauge theories.

The rest of the paper is organized as follows. In Sec.~\ref{origional-MMO}, we review the construction of the matrix model for a non-rotating gluon plasma in the infinite-volume limit and re-derive this model by using the Euclidean cylindrical coordinates that are suitable for generalization to the rotating case. In Sec.~\ref{MMO-in-real-rotation}, a detailed derivation of the matrix model is presented for a rotating system confined within a cylinder of finite radius. The rotation effect is introduced by computing the partition function in cylindrical coordinates that corotate with the system at a constant angular velocity.
In Sec.~\ref{phase-transition-in-real-rotation}, we analytically examine the properties of the perturbative and non-perturbative contributions that constitute the matrix model. In addition, based on the numerical evaluations, we discuss the effect of rotation on the deconfining temperature and the Polyakov loop, and also consider the inhomogeneity of the rotating system. The summary and outlooks are given in Sec.~\ref{summary}.

\section{Review of the original Matrix model}
\label{origional-MMO}

For $SU(N)$ pure gauge theories, the original matrix model for deconfinement was first proposed by considering the 1-loop effective potential in the presence of a classical background field, assuming gluon and ghost fields share the same effective mass $M$~\cite{Meisinger:2001cq}. This model consists of two kinds of contributions, corresponding to the first two terms in the high temperature expansion of the effective potential under the assumption $M\ll T$. Therefore, besides the usual perturbative contribution $\sim T^4$, a non-perturbative term $\sim M^2 T^2$ has been incorporated into the matrix model.
The background field $A_{0}^{\rm cl}$ is given by a diagonal matrix in color space,
\beq
(A_{0}^{\rm cl})_{ab}=\frac{2\pi T }{g} {\sf{q}}^a \delta_{ab}\, ,
\eeq
satisfying the traceless condition $\sum_{a=1}^{N} {\sf{q}}^a=0$ with $a,b=1,\cdots, N$ being the color indices in the fundamental representation. The gauge-invariant Polyakov loop is 
\beq
\ell=\frac{1}{N}{\rm Tr}\,{\cal P}\,{\rm exp}\Big(i g\int^{\beta}_0 A^{\rm cl}_{0}\, d\tau \Big)\, ,
\eeq
where ${\cal P}$ denotes the time ordering, $\tau$ is the imaginary time and $\beta\equiv1/T$ is the inverse temperature. In the presence of a background field, the gauge fields $A_{\mu}$ can be expanded as $A_{\mu}=A_{\mu}^{\rm cl}+B_{\mu}$ with $B_{\mu}$ being the quantum fluctuations and $A_{\mu}^{\rm cl}=A_{0}^{\rm cl} \delta_{\mu 0}$. The effective potential is then related to the logarithm of the partition function which is a functional integral involving the fluctuations of the gauge fields and the Faddeev-Popov ghosts.  
Dropping the interaction terms, the free Lagrangian density in Euclidean spacetime reads 
\beq
\label{lagrangian-1}
\mathcal{L}_0 ={\rm Tr} [ (D_{\mu}^{\rm cl} B_{\nu})^2- D_{\mu}^{\rm cl} B_{\nu} D_{\nu}^{\rm cl}  B_{\mu}] + \frac{1}{\xi} {\rm Tr} [ (D_{\mu}^{\rm cl}B_{\mu})^2]-2{\rm Tr}[\bar{\eta} (D_{\mu}^{\rm cl})^2\eta]\,, 
\eeq
where $D_{\mu}^{\rm cl}=\partial_{\mu}-i g [A_{\mu}^{\rm cl}, \quad]$ is the covariant derivative in the adjoint representation and $\xi$ is the gauge parameter. We also assume that the background field varies slowly and neglect the contribution from its derivatives with respect to spacetime. To compute the functional integrals in the partition function, we need to rewrite the above Lagrangian density as the following\footnote{It is valid under the integration over the spacetime.}  
\beq
\label{lagrangian}
\mathcal{L}_0 =-{\rm Tr} [B_{\mu} (D_{\nu}^{\rm cl})^2  B_{\mu}]+\big(1-\frac{1}{\xi}\big){\rm Tr} [B_{\nu} D_{\mu}^{\rm cl}D_{\nu}^{\rm cl}  B_{\mu}]-2{\rm Tr}[\bar{\eta} (D_{\mu}^{\rm cl})^2\eta]\,.
\eeq
In the above equation, the fluctuation operator $(D_{\nu}^{\rm cl})^2$ acts on either a scalar ghost field $\eta$ or a vector field $B_\mu$.
Furthermore, our calculations are performed in the double-line basis where bosonic fields are denoted as $B_{\mu}=B_{\mu}^{ba}t^{ab}$ and $\eta=\eta^{ba}t^{ab}$ with $t^{ab}$ being the generators of the $SU(N)$ gauge group and a pair of fundamental indices $(ab)$ referring to the color index in the adjoint representation. The trace of two generators defines the projection operator, ${\rm Tr}(t^{ab} t^{cd})=1/2 {\cal P}^{ab,cd}=1/2 {\cal P}^{ab}_{dc}$ with ${\cal P}^{ab}_{dc}=\delta^{a}_{d} \delta^{b}_{c}-1/N \delta^{ab} \delta_{dc}$. Notice that the covariant derivative in the adjoint representation becomes very simple as the commutator of the background field ${\sf q}$ with any generator $t^{ab}$ is just that generator multiplied by the difference of ${\sf q}$'s, i.e., $[{\sf{q}},t^{ab}] = ({\sf{q}}^a-{\sf{q}}^b)t^{ab} \equiv {\sf{q}}^{ab} t^{ab}$. More details of the double-line basis can be found in Refs.~\cite{Hidaka:2009hs,Cvitanovic:1976am}.

By introducing an effective mass for gluon and ghost fields, the effective Lagrangian density can be obtained as
\bqa
\label{eff-lagrangian}
\mathcal{L}^{\rm eff}_0 & =&-{\rm Tr} [B_{\mu} (D_{\nu}^{\rm cl})^2  B_{\mu}]-2{\rm Tr}[\bar{\eta} (D_{\nu}^{\rm cl})^2\eta]+ M^2 {\rm Tr} [ B_{\mu}B_{\mu} ] + 2 M^2 {\rm Tr}[\bar{\eta} \eta] \nonumber\\
& =& \frac{1}{2} {\cal P}^{ab}_{dc} \big\{ B_{\mu}^{dc} [ -(D_{\nu}^{{\rm cl},ab})^2 +M^2  ] B_{\mu}^{ba}+2\bar{\eta}^{dc}  [ -(D_{\nu}^{{\rm cl},ab})^2 +M^2  ] \eta^{ba} \big\}\, .
\eqa
In the second line of the above equation, we have taken the trace in color space and defined  $D_{\nu}^{{\rm cl},ab}=\partial_{\nu}-i (2\pi T) {\sf q}^{ab} \delta_{\nu0}$ acting on $B_\mu^{ba}$ or $\eta^{ba}$. For example, $D_{\nu}^{{\rm cl},ab}\eta^{ba}=\partial_{\nu}\eta^{ba}-i (2\pi T) {\sf q}^{ab} \eta^{ba}\delta_{\nu0}$ where $a$ and $b$ are fixed color indices, with no summation convention. From here on, we work in the Feynman gauge, $\xi=1$. Notice that Eq.~(\ref{lagrangian}) is obviously gauge invariant. Simply adding a mass term to the Lagrangian would violate the gauge invariance. Therefore, the resulting matrix model should only be considered as an illustration of a more self-consistent theory. 

Given the effective Lagrangian, the partition function can be obtained as 
\beq
\label{Z-old-1}
{\cal Z} = \int_{\rm periodic} [d B_{\mu}][d \eta] [d \bar{\eta}]{\exp}\bigg(\int_0^\beta d \tau \int d^3 x \,\mathcal{L}^{\rm eff}_0  \bigg) \,.
\eeq
After performing the functional integrals, we arrive at 
\beq
\label{Z-old-2}
{\cal Z}  =   {\cal N} \bigg \{  \prod_{\substack{a,b=1 \\ a \ne b}}^{N} [{\rm det} \,(\beta^2 \hat{{\cal O}}_{\rm v}^{ab} )]^{-1/2}{\rm det} \, (\beta^2\hat{{\cal O}}_{\rm s}^{ab})\bigg\} \bigg\{ [{\rm det} \,(\beta^2 \hat{{\cal O}}_{\rm v} )]^{-1/2}{\rm det} \, (\beta^2\hat{{\cal O}}_{\rm s})\bigg\}^{N-1}\,,
\eeq
where ${\cal N}$ is an irrelevant normalization constant. The first bracket in the above equation corresponds to the contributions from off-diagonal components of the bosonic fields $B_\mu^{ab}$ or $\eta^{ab}$($\bar{\eta}^{ab}$) with $a\neq b$, while the second bracket indicates that each of the $N-1$ diagonal components contributes equally to the partition function. To get the correct exponent $N-1$, we must replace the $N$ diagonal bosonic fields with only $N-1$ linearly independent combinations of $B_\mu^{aa}$ or $\eta^{aa}$($\bar{\eta}^{aa}$) because the diagonal generators satisfying $\sum_{a=1}^{N} t^{aa}=0$ are overcomplete. Since the diagonal gluon and ghost fields commute with the background field, their contributions don't depend on ${\sf q}$. If one is only interested in the $T$-dependent behavior of the background field or Polyakov loop, the second bracket can be dropped. However, we will keep these contributions to derive the complete matrix model. 

For a straightforward generalization to a rotating system, we make use of the Euclidean cylinder coordinates with $x^{\mu} = (\tau,\rho {\rm sin}\varphi, \rho {\rm cos}\varphi,z)$ to re-derive the original matrix model. Accordingly, the fluctuation operator acting on scalar ghost field is defined as $\hat{{\cal O}}_{\rm s}^{ab}=-(D_{\tau}^{{\rm cl},ab})^2- {\nabla}^2+M^2 \equiv - (D_{\rm s}^{{\rm cl},ab})^2  +M^2$ with ${\nabla}^2={\vec \nabla}\cdot {\vec \nabla}=\partial^2_{\rho}+\frac{1}{\rho} \partial_{\rho}+\frac{1}{\rho^2}\partial^2_{\varphi}+\partial^2_{z}$ in the cylinder coordinates\footnote{In Eq.~(\ref{Z-old-2}), we drop the color indices of the fluctuation operator when it acts on the diagonal components of the bosonic fields because $D_{\tau}^{{\rm cl},ab}$ is reduced to $\partial_\tau$ when $a=b$.}. Similarly, $\hat{{\cal O}}_{\rm v}^{ab}$ denotes the fluctuation operator for vector gluon field and takes the following form
\begin{align}
\label{vactor-Laplacian}
\hat{{\cal O}}_{\rm v}^{ab} =\begin{pmatrix}
-(D_{\rm s}^{{\rm cl},ab})^2+M^2 & 0 & 0 & 0 \\
0 & -(D_{\rm s}^{{\rm cl},ab})^2+M^2+
\frac{1}{\rho^2} & \frac{2}{\rho^2} \frac{\partial}{\partial \varphi} & 0 \\
0 & -\frac{2}{\rho^2} \frac{\partial}{\partial \varphi} &  -(D_{\rm s}^{{\rm cl},ab})^2+M^2+
\frac{1}{\rho^2} & 0 \\ 0 & 0 & 0 & -(D_{\rm s}^{{\rm cl},ab})^2+M^2
\end{pmatrix} \, .
\end{align}

Considering a diagonalizable operator $\hat{{\cal O}}$ with eigenvalue ${\cal O}$, we have  
\bqa
{\rm ln \, det}\,\hat{{\cal O}}&=&{\rm Tr \, ln} \,\hat{{\cal O}} =\int_0^{\beta}d \tau \int d^3 {\bf x} \,\langle\tau,{\bf x}|{\rm ln}\,\hat{{\cal O}}|\tau,{\bf x} \rangle\nonumber\\
&=& \sum_{\{\rm w\}} \int_0^{\beta}d \tau \int d^3 {\bf x}\, \langle\tau,{\bf x}| {\cal O}_{\{\rm w\}}\rangle \ln {\cal O}_{\{\rm w\}} \langle{\cal O}_{\{\rm w\}}|\tau,{\bf x} \rangle\, ,
\eqa
where $\{\rm w\}$ denotes a set of quantum numbers that specify a eigenmode $\langle\tau,{\bf x}| {\cal O}_{\{\rm w\}}\rangle$. The calculation of the logarithm of the partition function reduces to solving the eigenvalue problem of scalar and vector fluctuation operators, which gives the eigenvalues and corresponding eigenmodes for the ghost and gluon fields, respectively. These eigenvalues determine the energy spectrum of the system. For the scaler ghost field, the corresponding eigenequation reads\footnote{In the following, we use $\Phi$ and $\Psi$ to denote the eigenmodes $\langle\tau,{\bf x}| {\cal O}_{\{\rm w\}}\rangle$ for the ghost and gluon fields, respectively. They are functions of the imaginary time $\tau$ and cylinder coordinates $\rho$, $\varphi$ and $z$.}   
\beq
\label{eig-ghost-1}
\hat{{\cal O}}^{ab}_{\rm s} \Phi(\tau,\rho,\varphi,z)=(E^{ab})^2 \Phi(\tau,\rho,\varphi,z)\, .
\eeq
Solving Eq.~(\ref{eig-ghost-1}) with the periodic boundary conditions $\tau \sim \tau+\beta$ and $\varphi \sim \varphi+2\pi$ in the infinite-volume limit, we obtain the normalized eigenmodes of the ghost field, 
\beq
\label{ghost-wave-function}
 \Phi_{n{\cal J}}(\tau,\rho,\varphi,z)=e^{-i\omega_n \tau-i m \varphi-ip_z z} J_m(\rho p_{\rho})\frac{\sqrt{ p_{\rho} T}}{2\pi }\, .
\eeq
In the above equation, the Matsubara frequency $\omega_n=2\pi n T$ with $n \in \mathbb{Z}$ and ${\cal J}=(m,p_{\rho},p_z)$ is a cumulative index in which $m \in \mathbb{Z}$ is the angular quantum number, $p_{\rho} \geq 0$ is the continuous radial quantum number, and $p_z \in \mathbb{R}$ is the continuous longitudinal quantum number. In addition, $J_m(x)$ is the Bessel function of the first kind. 
The corresponding eigenvalues are found to be 
\beq
\label{eigenvalue}
(E^{ab}_{n{\cal J}})^2=p_{\rho}^2+p_z^2+M^2+(\omega_{n}^{ab})^2\,. 
\eeq 
The above eigenvalues are independent of the angular quantum number $m$, however, to make the notation compact, we still use the cumulative index ${\cal J}$ to denote the quantum numbers. On the other hand, these eigenvalues depend on the background field via the ${\sf q}$-modified Matsubara frequency $\omega_{n}^{ab}=\omega_{n} + 2\pi n T {\sf q}^{ab}$. As expected, the eigenvalues in Eq.~(\ref{eigenvalue}) are identical to those obtained in the Cartesian coordinates.

For the gluon field, solving the following eigenequation with the same boundary condition as the ghost field
\beq
\label{eig-gluon}
 \hat{{\cal O}}_{\rm v}^{ab} \Psi (\tau,\rho,\varphi,z) =(E^{ab})^2 \Psi(\tau,\rho,\varphi,z)\, ,
\eeq
the four normalized eigenmodes associated with different gluon polarizations are found to be  
\bqa
\label{gluon-wave-function}
\Psi_{n{\cal J}}^{(1,2)}(\tau,\rho,\varphi,z)&= & \Phi_{n{\cal J}}(\tau,\rho,\varphi,z)\varsigma^{(1,2)}\,,  \nonumber \\  
 \Psi_{n{\cal J}}^{(3)} (\tau,\rho,\varphi,z)&= &e^{-i\omega_n \tau-i m\varphi-ip_z z} J_{m+1}(\rho p_{\rho})\frac{\sqrt{p_{\rho}T}}{2\pi } \varsigma^{(3)}\,, \nonumber  \\
 \Psi_{n{\cal J}}^{(4)} (\tau,\rho,\varphi,z)&= &e^{-i\omega_n \tau-i m \varphi-ip_z z} J_{m-1}(\rho p_{\rho})\frac{\sqrt{p_{\rho}T}}{2\pi } \varsigma^{(4)}\,,
\eqa
with
\bqa
\varsigma^{(1)}&=&(1,0,0,0)^{T}\,,\quad\quad\quad\,\,
\varsigma^{(2)}=(0,0,0,1)^{T}\,,\nonumber\\
\varsigma^{(3)}&=&\frac{1}{\sqrt{2}} (0,1,i,0)^{T}\,,\quad\quad
\varsigma^{(4)}=\frac{1}{\sqrt{2}}(0,1,-i,0)^{T}\,.
\eqa
In addition, the eigenvalues in Eq.~(\ref{eig-gluon}) are identical for each of the four eigenmodes, and also the same as that for the ghost field given in Eq.~(\ref{eigenvalue}). Since contributions to ${\rm ln}{\cal Z}$ from the two unphysically polarized eigenmodes $\Psi_{n{\cal J}}^{(1)}$ and $\Psi_{n{\cal J}}^{(2)}$ are completely canceled by the corresponding ghost contributions, only the two physically polarized eigenmodes $\Psi_{n{\cal J}}^{(3)}$ and $\Psi_{n{\cal J}}^{(4)}$ contribute to ${\rm ln}{\cal Z}$ which can be expressed as
\bqa
\label{Z-1}
{\rm ln}{\cal Z} &=& -\frac{T}{2} \sum_{ab} \bigg(1-\frac{1}{N}\delta^{ab} \bigg)  \sum_n  \sum_{\cal J} \int_0^{\beta} d\tau \int_0^{\infty} \rho d\rho \int_0^{2\pi} d\varphi \int_{-\infty}^{\infty} dz \nonumber\\
&\times&  [J_{m+1}^2(\rho p_{\rho}) +J_{m-1}^2(\rho p_{\rho})]{\rm ln} \big[\beta^2 (E^{ab}_{n{\cal J}})^2  \big]\, ,
\eqa
where $\sum_{{\cal J}}=\frac{1}{(2\pi)^2}\sum_m \int_{-\infty}^{\infty} dp_z \int_0^{\infty} dp_{\rho} p_{\rho}$ represents the sum-integrals over all possible eigenmodes. Consequently, we can obtain the effective potential in the presence of a background field,
\bqa
\label{potential}
{\cal V}&=&\frac{T}{2} \sum_{ab} \bigg(1-\frac{1}{N}\delta^{ab} \bigg) \sum_n \sum_{{\cal J}}
 [J_{m+1}^2(\rho p_{\rho}) +J_{m-1}^2(\rho p_{\rho})] {\rm ln} \big[\beta^2 (E^{ab}_{n{\cal J}})^2  \big]\nonumber \\
 &=&\frac{T}{(2\pi)^2} \sum_{ab} \bigg(1-\frac{1}{N}\delta^{ab} \bigg) \sum_n \int_{-\infty}^{\infty} dp_z \int_0^{\infty} dp_{\rho} p_{\rho}
{\rm ln} \big[\beta^2 (E^{ab}_{n{\cal J}})^2  \big]\, .
\eqa
In the second line of the above equation, we have carried out the sum over $m$ by using the identity,  
\beq\label{id}
\sum_{m=-\infty}^{\infty} J_m^2(x)=1\, . 
\eeq
The above result shows that in the infinite-volume limit, the effective potential has no dependence on spacetime, therefore, the system is homogeneous. However, as we will show in the next section, the gluon plasma becomes inhomogeneous under rotation with a real-valued angular velocity $\Omega$ along the $z$-axis, which imposes a radial size constraint due to causality. By assuming $M \ll T$, it is straightforward to derive the first two terms in the high-temperature expansion of the effective potential ${\cal V}$, leading to the matrix model for deconfinement. After carrying out the sum-integrals, the matrix model is obtained as~\cite{Meisinger:2001cq},
\beq
\label{old-MMO}
\mathcal{F} = \frac{2\pi^2 T^4}{3} \sum_{ab} B_4({\sf{q}}^{ab})+\frac{\pi^2 T^4}{45}+\frac{M^2 T^2}{2} \sum_{ab} B_2({\sf{q}}^{ab}) -\frac{M^2 T^2}{12}\,,
\eeq 
where the Bernoulli polynomials $B_2(x)=x^2-x+\frac{1}{6}$ and  $B_4(x)=x^2(1-x)^2-\frac{1}{30}$ for $0 \leq x \leq 1$. For arbitrary values of $x$, the argument of the above Bernoulli polynomials should be understood as $x-[x]$ with $[x]$ the largest integer less than $x$, which is nothing but the modulo
function.

The background field $\sf{q}$ can be determined via the variational approach $\partial {\cal F}/ \partial \sf{q}=0$. However, with only the perturbative terms $\sim T^4$ in Eq.~(\ref{old-MMO}), $\sf{q}$ is found to be always zero and the system remains in the completely deconfined phase characterized by $\ell=1$. To drive the system to confinement, the nonperturbative contribution $\sim M^2T^2$ becomes very essential which must be added to the perturbative effective potential. Consequently, $\sf{q}$ is a nonzero constant at low temperatures, leading to $\ell=0$. Above the critical temperature, it becomes a $T$-dependent function which vanishes in the high temperature limit. Given such a background field that is self-consistently obtained from the equation of motion, Eq.~(\ref{old-MMO}) is able to model the deconfining phase transition in the sense that the correct $T$-dependent behavior of the Polyakov loop can be reproduced and the predictions on the thermodynamics also agree with the lattice QCD simulations.

Explicitly, we parameterize the background field as ${\bf q} =(\sf{q},-\sf{q})$ for $SU(2)$ and ${\bf q} =(\sf{q},0,-\sf{q})$ for $SU(3)$. It can be shown that in the confined phase, the respective values of $\sf{q}$ are $1/4$ and $1/3$. The background field becomes $T$-dependent in the deconfined phase and is given by
\beq
\label{q-T} 
{\sf{q}}^{N=2}_{\rm decon} = \frac{1-\sqrt{1-3 M^2/(2\pi^2 T^2)}}{4}\,  ,\quad {\rm and}\quad
{\sf{q}}^{N=3}_{\rm decon} =
\frac{1-\sqrt{1-2M^2/(\pi^2 T^2)}}{4} \,  .
\eeq
In addition, the only parameter $M$ can be determined by requiring the phase transition happens at the deconfining temperature $T_d$, in other word, the two minima of the effective potential in confined and deconfined phases become degenerate at the phase transition point. As a result, we find $M/T_d= \sqrt{6} \pi/3$ and $M/T_d= 2\sqrt{10} \pi/9$ for $SU(2)$ and $SU(3)$, respectively.

\section{The matrix model under real rotation }\label{MMO-in-real-rotation}

We consider a gluonic system rotating uniformly at a constant angular velocity $\Omega$ around a fixed $z$-axis\footnote{To be specific, we assume the rotation is counterclockwise and $\Omega>0$.}. 
The causality requires that the local velocity must be smaller than the speed of light, and thus the product of the angular velocity
and the radius of the rigidly rotating system is bounded. Accordingly, we confine the gluon plasma within a cylinder of radius $\mathcal{R}$ and impose the constraint $\mathcal{R}\Omega<1$.
In order to effectively include the effect of rotation, we introduce a reference frame which corotates with the gluon plasma. The Euclidean cylinder coordinate in the corotating frame is denoted as $\tilde{x}^{\mu}=(\tilde{\tau}, \tilde{\rho}, \tilde{\varphi},\tilde{z})$. The tilded coordinates are related to those in the laboratory reference frame via the following transformations 
\beq
\label{coordinate_relation}
\tilde{\tau}=\tau\, , \quad \tilde{\varphi}=[\varphi-i \Omega \tau]_{2\pi}\, , \quad \tilde{\rho}=\rho\,, \quad \tilde{z}=z\, ,
\eeq
where $[\dots]_{2\pi}$ means modulo $2\pi$. In terms of the tilded coordinates, we can obtain the effective Lagrangian density in the corotating reference frame\footnote{This is equivalent to the Lagrangian in a curved spacetime which is obtained by using the curvilinear metric in the corotating frame.}. The result turns out to be very simple. As compared to its counterpart in the laboratory frame as given by Eq.~(\ref{eff-lagrangian}), only the temporal component of the covariant derivative gets modified, 
\bqa
\label{Ducl}
D_{\tilde{\tau}}^{\rm cl} =  \partial_{\tilde{\tau}}-i\Omega \partial_{\tilde{\varphi}} -i g [A_{0}^{\rm cl}, \quad]\, ,
\eqa
and accordingly,
\bqa
\label{Ducl}
D_{\tilde{\tau}}^{{\rm cl},ab} =  \partial_{\tilde{\tau}}-i\Omega \partial_{\tilde{\varphi}} -i(2\pi T){\sf q}^{ab}\, ,
\eqa
which explicitly depends on the angular velocity $\Omega$.
As a result, the partition function in the corotating frame takes a form similar to Eq.~(\ref{Z-old-2}), where  $\hat{{\cal O}}_{\rm s}^{ab}$ and $\hat{{\cal O}}_{\rm v}^{ab}$ should be replaced by $\hat{\tilde{{\cal O}}}_{\rm s}^{ab}$ and $\hat{\tilde{{\cal O}}}_{\rm v}^{ab}$, respectively. The fluctuation operator for scalar field $\hat{\tilde{{\cal O}}}_{\rm s}^{ab}=-(D_{\tilde{\tau}}^{{\rm cl},ab})^2-\tilde{\nabla}^2 +M^2$ and the Laplacian operator $\tilde{\nabla}^2$ is defined the same as before but all the coordinates are tilded. In addition, the fluctuation operator for vector field $\hat{\tilde{{\cal O}}}_{\rm v}^{ab}$ can be obtained from the tilded version of Eq.~(\ref{vactor-Laplacian}) in which one also changes $(D_{\rm s}^{{\rm cl},ab})^2$ into $(D_{\tilde{\tau}}^{{\rm cl},ab})^2+\tilde{\nabla}^2$.

The energy eigenequation of ghost field can be written as 
\bqa
\label{eigenequation-ghost}
\hat{\tilde{{\cal O}}}_{\rm s}^{ab} \Phi(\tilde{\tau}, \tilde{\varphi},\tilde{\rho}, \tilde{z}) & =& (\tilde{E}^{ab})^2\Phi (\tilde{\tau}, \tilde{\varphi},\tilde{\rho}, \tilde{z}) ,
\eqa
which is solved with the periodicity of the imaginary-time direction $\tilde{\tau} \sim \tilde{\tau}+\beta$ and the Dirichlet boundary condition $\Phi (\tilde{\tau}, \tilde{\varphi},\mathcal{R}, \tilde{z})=0$. Notice that the continuous radial quantum number $p_{\rho}$ in the infinite-volume limit is now discretized due to the boundary condition, therefore, the cumulative index ${\cal J}=(m,l,p_z)$ with an integer quantum number $l \geq  1$. Furthermore, the normalized eigenmode is found to be
\bqa
\label{ghost-wave-function-w}
\Phi_{n{\cal J}} (\tilde{\tau},\tilde{\rho},\tilde{\varphi},\tilde{z})=e^{-i\omega_n \tilde{\tau}-i m \tilde{\varphi}-ip_z \tilde{z}}  \frac{J_m (\frac{\tilde{\rho}}{\mathcal{R}} \kappa_{{m}, l})}{\sqrt{2\beta}\pi \mathcal{R}  |J_{m+1}(\kappa_{{m}, l}) | }, 
\eqa
where the dimensionless quantity $\kappa_{{m}, l}$ is the $l$th positive root of the Bessel function of the first kind, determined by $J_m(\kappa_{{m} ,l})=0$.  
The associated eigenvalue in the corotating frame reads 
\bqa
(\tilde{E}^{ab}_{n{\cal J}})^2=p_z^2+\bigg(\frac{\kappa_{{m}, l}}{\mathcal{R}}\bigg)^2 +M^2 +(\omega_{n}^{ab}- i m \Omega )^2\,.
\eqa 
It demonstrates that similar to the background field which acts as an imaginary chemical potential, the rotation can be incorporated through a shift of the Matsubara frequency, and thus plays the role as a real chemical potential. 

For the vector gluon fields, solving the following eigenequation with the same boundary conditions as ghost 
\bqa
\label{eig-ghost}
\hat{\tilde{{\cal O}}}_{\rm v}^{ab} \Psi(\tilde{\tau}, \tilde{\varphi},\tilde{\rho}, \tilde{z})=(\tilde{E}^{ab})^2 \Psi(\tilde{\tau}, \tilde{\varphi},\tilde{\rho}, \tilde{z})\,,
\eqa
we obtain four normalized eigenmodes, 
\bqa
\label{gluon12-wave-function-w}
& & \Psi_{n{\cal J}}^{(1,2)}(\tilde{\tau}, \tilde{\varphi},\tilde{\rho}, \tilde{z})=\Phi_{n{\cal J}} (\tilde{\tau}, \tilde{\varphi},\tilde{\rho}, \tilde{z}) \varsigma^{(1,2)},  \\ 
\label{gluon3-wave-function-w}
& & \Psi_{n{\cal J}}^{(3)}(\tilde{\tau}, \tilde{\varphi},\tilde{\rho}, \tilde{z})= e^{-i\omega_n \tilde{\tau}-i m \tilde{\varphi}-ip_z \tilde{z}}  \frac{J_{m+1} (\frac{\tilde{\rho}}{\mathcal{R}} \kappa_{{m+1},l})}{\sqrt{2\beta}\pi \mathcal{R}  |J_{m+2}(\kappa_{{m+1},l}) | } \varsigma^{(3)},  \\
\label{gluon4-wave-function-w}
& & \Psi_{n{\cal J}}^{(4)}(\tilde{\tau}, \tilde{\varphi},\tilde{\rho}, \tilde{z})= e^{-i\omega_n \tilde{\tau}-i m \tilde{\varphi}-ip_z \tilde{z}}  \frac{J_{m-1} (\frac{\tilde{\rho}}{\mathcal{R}} \kappa_{{m-1},l})}{\sqrt{2\beta}\pi \mathcal{R}  |J_{m}(\kappa_{{m-1},l}) | } \varsigma^{(4)}\,,
\eqa 
together with the corresponding eigenvalues 
\bqa
&& (\tilde{E}^{ab,(1,2)}_{n{\cal J}})^{2}=p_z^2+\bigg(\frac{\kappa_{{m}, l}}{\mathcal{R}}\bigg)^2 +M^2 +(\omega_{n}^{ab}- i m \Omega )^2\,, \nonumber \\
&& (\tilde{E}^{ab,(3)}_{n{\cal J}})^{2}=p_z^2+\bigg(\frac{\kappa_{{m+1},l}}{\mathcal{R}}\bigg)^2 +M^2 +(\omega_{n}^{ab}- i m \Omega )^2\,, \nonumber \\
&& (\tilde{E}^{ab,(4)}_{n{\cal J}})^{2}=p_z^2+\bigg(\frac{\kappa_{{m-1},l}}{\mathcal{R}}\bigg)^2 +M^2+(\omega_{n}^{ab}- i m \Omega )^2\,.  
\eqa

Given the above results, it is straightforward to derive the logarithm of the partition function for the gluonic system under real rotation. Just like before, only the physically polarized eigenmodes contribute and the result can be written as
\bqa
\label{lnz-w}
{\rm ln}{\cal Z}({\Omega}) &=& -\frac{T}{2}\sum_{ab} \bigg(1-\frac{1}{N}\delta^{ab} \bigg)  \sum_n  \sum_{{\cal J}} \int_0^{\beta} d \Tilde{\tau} \int_0^{\cal R} {\tilde{\rho}} d {\tilde{\rho}} \int_0^{2\pi} d{\tilde{\varphi}} \int_{-\infty}^{\infty} d{\tilde{z}} \nonumber\\
&\times&  \bigg\{  {\rm ln} \big[\beta^2 (\tilde{E}^{ab,(3)}_{n{\cal J}})^{2}  \big]  \frac{J_{m+1}^2 (\frac{\tilde{\rho}}{\mathcal{R}} \kappa_{{m+1},l})}{J_{m+2}^2(\kappa_{{m+1},l})}  +{\rm ln} \big[\beta^2 (\tilde{E}^{ab,(4)}_{n{\cal J}})^{2}  \big]  \frac{J_{m-1}^2 (\frac{\tilde{\rho}}{\mathcal{R}} \kappa_{{m-1},l})}{J_{m}^2(\kappa_{{m-1},l})} \bigg\},
\eqa
with $\sum_{{\cal J}}= \frac{1}{2\pi^2 {\mathcal{R}}^2} \sum_{m=-\infty}^{\infty} \sum_{\ell=1}^{\infty} \int_{-\infty}^{\infty} d p_z $ being the sum-integrals over all possible eigenmodes. Consequently, the effective potential can be expressed as 
\bqa
\label{potential-w-rho}
{\cal V}(\Omega,\tilde{\rho}) & =&\frac{T}{2} \sum_{ab} \bigg(1-\frac{1}{N}\delta^{ab} \bigg) \sum_n \sum_{{\cal J}}
\bigg\{  {\rm ln} \big[\beta^2 (\tilde{E}^{ab,(3)}_{n{\cal J}})^{2}  \big]  \frac{J_{m+1}^2 (\frac{\tilde{\rho}}{\mathcal{R}} \kappa_{{m+1},l})}{J_{m+2}^2(\kappa_{{m+1},l})} \nonumber \\ &+& {\rm ln} \big[\beta^2 (\tilde{E}^{ab,(4)}_{n{\cal J}})^{2}  \big]  \frac{J_{m-1}^2 (\frac{\tilde{\rho}}{\mathcal{R}} \kappa_{{m-1},l})}{J_{m}^2(\kappa_{{m-1},l})}           \bigg\}\, ,
\eqa
which is obviously inhomogeneous along the radial direction due to the dependence on $\tilde{\rho}$. Such an inhomogeneity arises in a cylinder of finite radius, regardless of whether the rotation effect is taken into account. On the other hand, by setting $\Omega=0$ and then taking the limit ${\cal R}\rightarrow \infty$ in the above equation, we can reproduce the result as given in Eq.~(\ref{potential}). Notice that as ${\cal R}\rightarrow \infty$, the summation over the radial quantum number $l$ receives dominant contributions from large $l$, so that $\kappa_{m,l}\sim {\cal O}({\cal R})$. From the asymptotic form of Bessel functions, we can replace the sum over $l$ by the integral $({\cal R}/\pi)\int_0^\infty d p_\rho$ with $p_\rho=\kappa_{m,l}/R$. In addition, for the squared Bessel functions in the normalization factor, we have $J_{m+1}^2(\kappa_{m,l})\rightarrow2/(\pi {\cal R}p_\rho)$. As a result, summing over $m$ using Eq.~(\ref{id}) yields a homogeneous effective potential identical to Eq.~(\ref{potential}). This consistency check also indicates that if we take the effective mass $M$ in the above equation as a constant, it must be identical to that in the infinite-volume limit. 

To proceed, we analytically perform the Matsubara frequency sum in Eq.~(\ref{potential-w-rho}). After dropping the zero-point energy, the effective potential becomes
\bqa
\label{potential-w-rho-n}
{\cal V} (\Omega,\tilde{\rho})&=& \frac{T}{2} \sum_{ab} \bigg(1-\frac{1}{N}\delta^{ab} \bigg) \sum_{{\cal J}}
\bigg\{\sum_{\sigma=\pm}  {\rm ln} \big[1-e^{-\beta \epsilon^{ab,(3)}_{\sigma, \cal J}(\Omega)} \big]  
\frac{J_{m+1}^2 (\frac{\tilde{\rho}}{\mathcal{R}} \kappa_{{m+1},l})}{J_{m+2}^2(\kappa_{{m+1},l})} 
\nonumber \\ &+ & \sum_{\sigma=\pm}   {\rm ln} \big[1-e^{-\beta \epsilon^{ab,(4)}_{\sigma, \cal J}(\Omega)} \big]  \frac{J_{m-1}^2 (\frac{\tilde{\rho}}{\mathcal{R}} \kappa_{{m-1},l})}{J_{m}^2(\kappa_{{m-1},l})} \bigg\}\, ,
\eqa
where we define the following effective energies
\bqa
\epsilon^{ab,(3)}_{\pm, \cal J}(\Omega)&=&\sqrt{p_z^2+(\kappa_{{m+ 1},l}/{\mathcal{R}})^2 +M^2}\pm m \Omega\pm i 2\pi T {\rm q}^{ab}\, ,\nonumber \\
\epsilon^{ab,(4)}_{\pm,\cal J}(\Omega)&=&\sqrt{p_z^2+(\kappa_{{m- 1},l}/{\mathcal{R}})^2 +M^2}\pm m \Omega\pm i 2\pi T {\rm q}^{ab}\, .
\eqa
For diagonal contributions with $a=b$, using the properties of the zeros of Bessel functions $\kappa_{{m},l}\ge \kappa_{{m},1}$ and $\kappa_{{m},1}>1+|m|$, it can be shown that the effective energies are always positive due to the constraint of causality ${\cal R}\Omega<1$. For off-diagonal contributions $a\neq b$, they become complex valued. However, the effective potential is still real after summing over the colors. To get the matrix model under real rotation, we can expand the above effective potential at high temperatures, and the first two terms in the expansion are given by
\bqa
\label{potential-w-dV}
{\cal F}(\Omega, \tilde{\rho}) & =& \frac{M^2}{2\pi^2 {\mathcal{R}}^2} \sum_{s,l=1}^{\infty} \sum_{m=-\infty}^{\infty} 
\sum_{ab} \bigg( 1-\frac{1}{N}\delta^{ab} \bigg)\bigg[ K_0 \bigg(\beta s \frac{\kappa_{{m+1},l}}{\mathcal{R}} \bigg)    -\frac{2\kappa_{{m+1},l}}{s\beta M^2\mathcal{R}} K_1 \bigg(\beta s \frac{\kappa_{{m+1},l}}{\mathcal{R}} \bigg)\bigg]\nonumber \\&\times& {\rm cos}(2 s\pi {\rm{q}}^{ab})  \big[ {\rm cosh}(s \beta  m  \Omega) + {\rm cosh}(s \beta (m + 2) \Omega)
\big] \frac{J_{m+1}^2 (\frac{\tilde{\rho}}{\mathcal{R}} \kappa_{{m+1},l})}{J_{m+2}^2(\kappa_{{m+1},l})}
\,,
\eqa
where $K_0(x)$ and $K_1(x)$ are the modified Bessel functions of the second kind. Furthermore, using the identity 
\bqa
 \int_0^{\cal R} \tilde{\rho}  \frac{J_{m}^2 (\frac{\tilde{\rho}}{\mathcal{R}} \kappa_{{m},l})}{J_{m+1}^2(\kappa_{{m},l})} d \tilde{\rho} =\frac{{\cal R}^2}{ 2}, 
\eqa
we can perform the integrals over spacetime in Eq.~(\ref{lnz-w}) and define the bulk-averaged effective potential ${\bar {\cal V}}(\Omega)={\rm ln}{\cal Z}(\Omega)/(\beta V)$. Accordingly, the bulk-averaged matrix model can be obtained as
\bqa
\label{potential-w-aver}
{\bar {\cal F}}(\Omega) & =& \frac{M^2}{2\pi^2 {\mathcal{R}}^2} \sum_{s,l=1}^{\infty} \sum_{m=-\infty}^{\infty} 
\sum_{ab} \bigg( 1-\frac{1}{N}\delta^{ab} \bigg)\bigg[ K_0 \bigg(\beta s \frac{\kappa_{{m+1},l}}{\mathcal{R}} \bigg)    -\frac{2\kappa_{{m+1},l}}{s\beta M^2\mathcal{R}} K_1 \bigg(\beta s \frac{\kappa_{{m+1},l}}{\mathcal{R}} \bigg)\bigg]\nonumber \\&\times& {\rm cos}(2 s\pi {\rm{q}}^{ab}) \big[{\rm cosh}(s \beta  m  \Omega) + {\rm cosh}(s \beta (m + 2) \Omega)
\big] \,.
\eqa
In Eqs.~(\ref{potential-w-dV}) and (\ref{potential-w-aver}), terms associated with $K_1(x)$ correspond to the perturbative contribution in the matrix model which is independent of the effective mass $M$, while terms associated with $K_0(x)$ represent the corresponding non-perturbative contribution in this model which is proportional to $M^2$.

\section{The deconfining phase transition under real rotation} \label{phase-transition-in-real-rotation}

In this section, we discuss the deconfining phase transition for $SU(2)$ and $SU(3)$ gauge theories by employing the matrix model under real rotation as given by Eq.~(\ref{potential-w-dV}) or ({\ref{potential-w-aver}}). For convenience, we set the cylinder radius $\mathcal{R}=1/T_d$. According to the deconfining temperatures\footnote{$T_d$ is the phase transition temperature for a non-rotating system in the infinite-volume limit, while the phase transition temperature under rotation is denoted as $T_c$ in the following.} from lattice simulations~\cite{Lucini:2012gg}, the values of $\mathcal{R}$ are $\sim 0.73 \,{\rm fm}$ and $\sim 0.67\,{\rm fm}$ for $SU(3)$ and $SU(2)$, respectively. Although relatively small, they are still comparable to the typical size of the QGP produced in the non-central collisions. We can expect that the small cylinder radius considered here will result in a significant finite-size effect. On the other hand, choosing large values of ${\cal R}$ leads to a slow convergence of the sums in the effective potential which leads to a rather time-consuming evaluation. We further note that when scaled by a factor of $1/T_d^4$, both $\mathcal{F}$ and $\bar{\mathcal{F}}$ can be expressed in terms of dimensionless variables, including $\Omega\mathcal{R}$, $T/T_d$ and $\tilde{\rho}/\mathcal{R}$, while for the dimensionless quantity $M/T_d$, we consider two cases in our numerical evaluations. In Sec.~\ref{mconstant}, we assume $M/T_d$ to be a constant, therefore, it remains those values given in Sec.~\ref{origional-MMO}. A possible $\Omega$-dependence of $M/T_d$ is discussed in Sec.~\ref{mgw} where $M$ is assumed to depend on an $\Omega$-dependent running coupling.

\subsection{The case without the non-perturbative contribution, $M=0$}\label{sec-Mg0}

The perturbative contribution describes the high-temperature behavior of the matrix model, and can be systematically improved by computing the high-order effective potential~\cite{Dumitru:2013xna,Guo:2018scp,KorthalsAltes:1993ca}. It has been shown that for pure imaginary rotations, a deconfining phase transition can occur perturbatively~\cite{Chen:2022smf,Zhang:2026ctc}. 
On the other hand, for a real-valued angular velocity, the minimum of the perturbative effective potential is located at ${\sf q}=0$, or the equivalent vacua that appear periodically. As a result, the system will always stay in the completely deconfined phase and no deconfining phase transition could happen. Notice that the ${\sf q}$-dependence in Eqs.~(\ref{potential-w-dV}) and ({\ref{potential-w-aver}}) is simply given by ${\rm cos}(2 s \pi {\sf q}^{ab})$ and the modified Bessel function $K_1(x)>0$, consequently, the perturbative effective potential equals the cosine function multiplied by a negative factor, and thus its minima are attained at ${\rm cos}(2 s \pi {\sf q}^{ab})=1$. This result holds for any $\tilde{\rho}$ and generalizes the finding in Ref.~\cite{Chen:2022smf} where the authors have numerically shown that there is no phase transition along the rotation axis at $\tilde{\rho}=0$.

\begin{figure}[htbp]
    \centering
\includegraphics[width=0.45\textwidth]{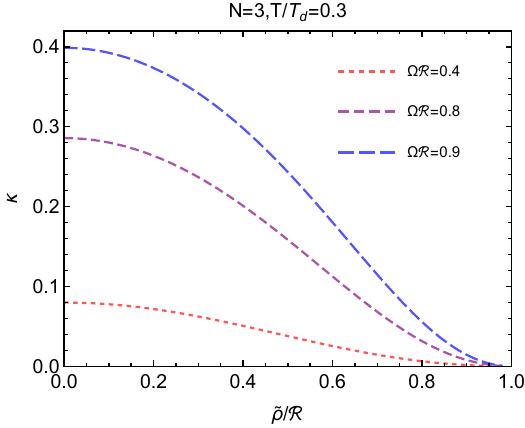}
\hfill 
\includegraphics[width=0.45\textwidth]{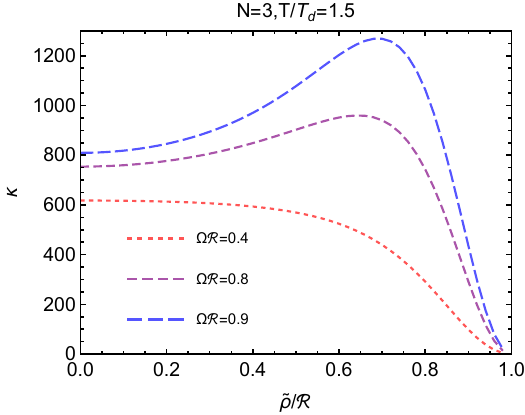}
    \caption{The curvature of the perturbative vacuum as a function of $\Tilde{\rho}/\mathcal{R}$ at $T/T_d=0.3$ (left panel) and at $T/T_d=1.5$ (right panel). }
    \label{curer}
\end{figure}
The numerical results in Ref.~\cite{Chen:2022smf} also demonstrated that, along the rotation axis, the perturbative vacuum becomes more stable with increasing angular velocity as it increases the curvature of this vaccum. The curvature $\kappa$ is defined as $\kappa=|\mathcal{F}''|/[1+(\mathcal{F}')^2]^{3/2}$ with $\mathcal{F}'=\partial \mathcal{F}/\partial {\sf q}$. For vanishing $\mathcal{F}'$ at the minimum, it reduces to $\kappa=|\mathcal{F}''|$. According to Eq.~(\ref{potential-w-dV}), the $\Omega$-dependence of the curvature only lies in the hyperbolic cosine function, i.e., $\kappa\sim {\rm cosh}(s\beta m \Omega)$. For a given temperature, we can therefore prove that, regardless of the radial position, $\kappa$ always increases with $\Omega$ and enhanced stability can be expected at larger angular velocity. In addition, the curvature also increases with temperature $T$, as can be proved by showing that, for $a>b$, the function $K_1(a/T) \cosh(b/T)$ increases with $T$\footnote{According to the explicit form of $a$ and $b$ in question, $a=s\kappa_{m,l}/\mathcal{R} $ and $b=s (m\pm1) \Omega$, the condition $a>b$ can be satisfied because of the constraint of causality.}. 
Therefore, the perturbative vacuum becomes more stable at higher temperatures.

The gluonic system under real rotation becomes spatially inhomogeneous, therefore, the stability of the perturbative vacuum also varies with the radial position. The corresponding $\tilde{\rho}$-dependence of the curvature lies in the rapidly oscillating Bessel function $J_m(x)$, according to Eq.~(\ref{potential-w-dV}). For relatively low temperatures, the stability of the vacuum decreases monotonically with increasing $\tilde{\rho}$. However, a non-monotonic behavior shows up when both the temperature and the angular velocity become large. In this case, $\kappa$ increases away from the rotation axis, then it drops down at some intermediate radial position $\tilde{\rho}$, featuring a local maximum stability of the system. These behaviors are illustrated in Fig.~\ref{curer} where we plot $\kappa$ as a function of $\Tilde{\rho}$ for three different angular velocities at $T/T_d=0.3$ and $T/T_d=1.5$ for $SU(3)$. We also find that, with a further increase in temperature, this non-monotonic behavior emerges earlier as $\Omega$ is gradually increased. These conclusions are also applicable to $SU(2)$.

\subsection{The case with the non-perturbative contribution, $M \neq 0$ }\label{mg}

Since the perturbative effective potential is insufficient to drive the system to a confined phase, adding a non-perturbative contribution becomes necessary to study the phase transition of the gluonic system under real rotation. Recall that as compared to a non-rotating system in the infinite-volume limit, both the real rotation and the finite-volume configuration don't move the locations of the perturbative minima, therefore, it would be interesting to know if the same also holds for the non-perturbative contribution in the matrix model. It turns out that the non-perturbative contribution $\sim M^2$ in Eq.~(\ref{potential-w-dV}) favors a confined phase with a vanishing Polyakov loop, indicating its minimum is always located at ${\sf q}= 1/4$ for $SU(2)$ and ${\sf q} =1/3$ for $SU(3)$, which is the same as the non-rotating system in the infinite-volume limit. This can be proved by investigating the first derivative of the non-perturbative contribution with respect to the background field ${\sf{q}}$. Using the following integral representation of the modified Bessel function
\beq
K_0(x)=\int_0^\infty e^{-x \cosh \mu} d \mu\,,
\eeq
and taking into account the causality constraint, we can carry out the sum over $s$ and show that this derivative can be expressed by an identically negative term multiplied by $\sin(4\pi {\sf q})$ and $\sin(2\pi {\sf q})+\sin(4\pi {\sf q})$ for $SU(2)$ and $SU(3)$, respectively. Furthermore, since the non-perturbative contribution is a periodic function of ${\sf q}$, and in the case of $SU(2)$, it is invariant under ${\sf q}\rightarrow 1/2-{\sf q}$ when $0<{\sf q}<1/2$, we can only focus on the interval of $0<{\sf q}<1/4$. Clearly, the derivative is always negative in this interval, indicating that the non-perturbative contribution decreases monotonically with increasing ${\sf q}$ and reaches its minimum at ${\sf q}=1/4$. Similarly, we can consider the interval of $0<{\sf q}<1/2$ for $SU(3)$. The derivative is negative if $0<{\sf q}<1/3$, then it becomes positive when $1/3<{\sf q}<1/2$. Therefore, ${\sf q}=1/3$ corresponds to the minimum in the case of $SU(3)$.

According to the above discussions, the locations of the minima for the perturbative or non-perturbative contribution alone are not changed when considering a rotating gluon plasma in a cylindrical volume. As a result, we can expect that the phase transition is induced by a mechanism similar to that of the non-rotating system in the infinite-volume limit. For example, at a fixed position in the cylinder, the vacuum at low temperatures corresponds to ${\sf q}=1/3$ when $SU(3)$ is considered, whereas at high temperatures it is determined by a background field with a highly nontrivial dependence on both $T$ and $\Omega$. The phase transition happens when the two vacua become degenerate. In Sec.~\ref{origional-MMO}, we show that such a degeneracy requires $T_d= 9\sqrt{10} M /(20 \pi)$. For a rotating system, the deconfining temperature $T_c$ is determined in a similar way by equating the low-temperature vacuum to that at high temperatures; this procedure, however, must be carried out numerically. In the following, we study the deconfining phase transition based on the matrix model and illustrate how the deconfining temperature $T_c$ and the Polyakov loop $\ell$ change when varying the angular velocity $\Omega$ as well as the radial position ${\tilde \rho}$.

\subsubsection{Phase transition with a constant $M$}\label{mconstant}

\begin{figure}[htbp]
    \centering    \includegraphics[width=0.45\textwidth]{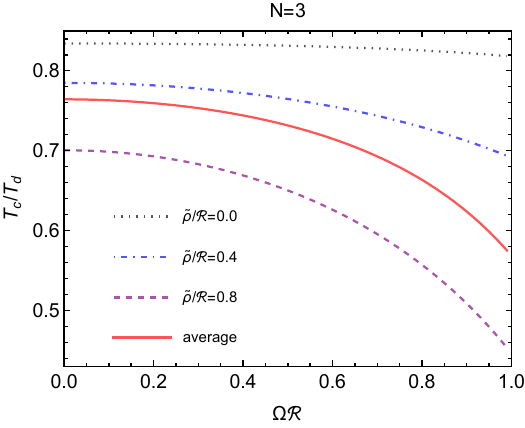} 
    \hfill 
\includegraphics[width=0.45\textwidth]{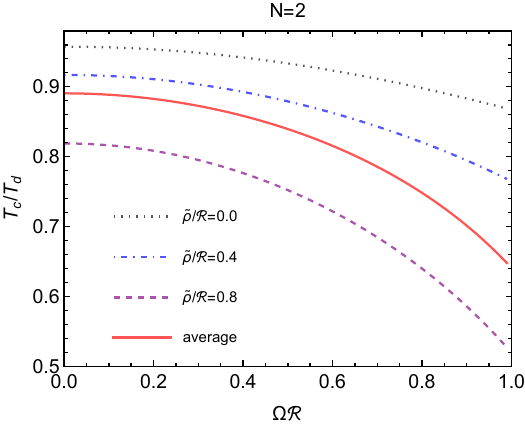} 
\caption{The temperature ratio $T_c/T_d$ as a function of $\Omega \mathcal{R}$.}
    \label{TTdrTdvalueRe}
\end{figure}

We first consider the phase transition with a constant $M$. In terms of the previously introduced dimensionless variables, for the given position ${\tilde \rho}/{{\cal R}}$ and angular velocity $\Omega {{\cal R}}$, we determine the temperature ratio $T_c/T_d$ which actually reflects the change of the deconfining temperature $T_c$ as compared to that for a non-rotating system in the infinite-volume limit. Fig.~\ref{TTdrTdvalueRe} shows $T_c/T_d$ as a function of $\Omega \mathcal{R}$ for different radial positions. For both $SU(2)$ and $SU(3)$, the temperature ratio is always less than unity. Furthermore, $T_c$ decreases with increasing $\Omega$ and for relatively small $\Omega$, such a behavior can be perfectly described by the following quadratic fit,
\beq
\label{fit2}
T_c(\Omega)=T_c(\Omega=0)-{\rm c}\, \Omega^2 \,, \quad{\rm with}\quad\Omega \mathcal{R}\lesssim 0.5 \,.
\eeq
The above equation holds for any fixed $\tilde{\rho}$ and ${\rm c}$ denotes a positive number whose value only depends on $\tilde{\rho}$.  
The decrease in $T_c$ becomes rather rapid when both $\Omega$ and ${\tilde \rho}$ get very large which can be understood as a consequence of the centrifugal effects induced by the real rotation. These findings indicate that real rotation favors deconfinement as it accelerates the transition to the deconfined phase. Nevertheless, no definitive conclusion can be drawn at present. Although some other models, such as the bag model~\cite{Mameda:2023sst}, hadron resonance gas model~\cite{Fujimoto:2021xix}, analytically tractable model in compact QED~\cite{Chernodub:2020qah} as well as the caloron model~\cite{Jiang:2024zsw} predict a decreased $T_c$ in a rotating system, most lattice simulations~\cite{Braguta:2020biu, Braguta:2021jgn,Braguta:2021ucr,Braguta:2023iyx} support the opposite conclusion, namely, $T_c$ increases with increasing $\Omega$ and real rotation favors confinement\footnote{The $\Omega$-dependence of $T_c$ found in lattice simulations also shows a quadratic behavior.}. However, by using strong coupling expansion method, the result from recent lattice simulations turns to be consistent with our conclusion~\cite{Wang:2025mmv}. Note that earlier lattice simulations did not take into account the $\tilde{\rho}$-dependence of the deconfining temperature, and their results can therefore be understood as being bulk-averaged. We also present the corresponding result in Fig.~\ref{TTdrTdvalueRe} by the solid line which also exhibits the same behavior as those $\tilde{\rho}$-dependent ones. In addition, there are also relevant studies based on holographic QCD and the obtained results~\cite{Chen:2020ath,Zhao:2022uxc,Braga:2022yfe,Wang:2024szr,Chen:2024jet,Chen:2024edy,Li:2026pkd} with different models still don't converge, indicating the necessity of further investigations on this issue.

In addition, we want to point out that although the deconfining temperature along the rotation axis is almost unaffected by rotation for $SU(3)$, a significant decrease of $T_c$ with increasing $\Omega$ is observed at ${\tilde \rho}=0$ in the case of $SU(2)$. The same behavior in $SU(3)$ has also been observed in lattice simulations which, however, don't provide the corresponding result for $SU(2)$~\cite{Braguta:2023iyx}. On the other hand, the caloron model~\cite{Jiang:2023zzu} shows a similar decrease of $T_c$ along the rotation axis for $SU(2)$, provided that the coupling constant doesn't depend on the angular velocity. It is not clear what causes such a discrepancy between $SU(3)$ and $SU(2)$, but we note that the fundamental difference, namely, that the phase transition is first order for $SU(3)$ and second order for $SU(2)$, still holds true in a rotating plasma.
 
\begin{figure}[htbp]
    \centering
\includegraphics[width=0.45\textwidth]{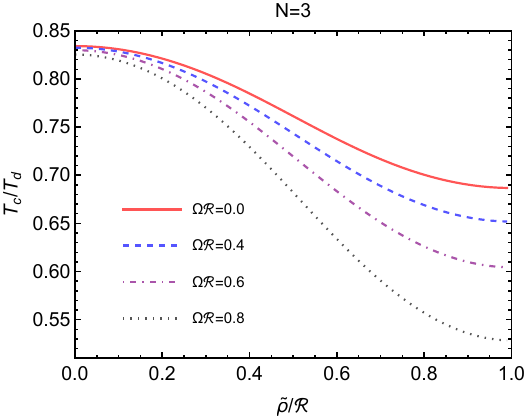}
\hfill 
\includegraphics[width=0.45\textwidth]{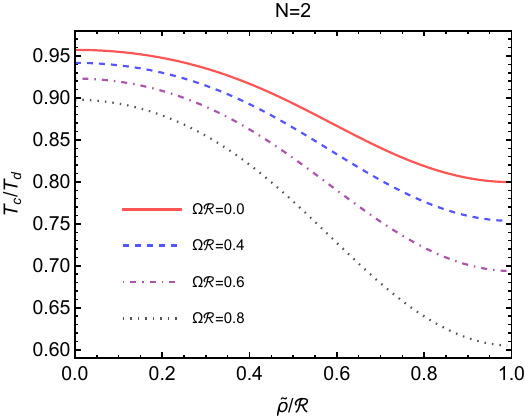}
    \caption{The temperature ratio $T_c/T_d$ as a function of $\tilde{\rho}/\mathcal{R}$.}
    \label{TTdwTdvalueRe}
\end{figure}

In Fig.~\ref{TTdwTdvalueRe}, we present $T_c/T_d$ as a function of $\tilde{\rho}/\mathcal{R}$ for different angular velocities. It shows a decreased $T_c$ with increasing $\tilde{\rho}$ for both $SU(2)$ and $SU(3)$, indicating the phase transition occurs more readily at positions far from the rotation axis which is consistent with the Tolman-Ehrenfest law~\cite{Tolman:1930ona}, i.e., real rotation effectively heats the system outside of the rotation axis. As a result, the rotating gluonic system exhibits not only the confined and deconfined phases, but also a mixed phase. For a given angular velocity, when the temperature is larger than the deconfining temperature along the rotation axis which is denoted as $T_c(\tilde{\rho}= 0)$, the system is entirely in the deconfined phase. Accordingly, a fully confined phase can be achieved when the temperature is lower than $T_c(\tilde{\rho}\rightarrow{\cal R})$\footnote{We do not consider $\tilde{\rho}=\mathcal{R}$ since the effective potential vanishes there due to the Dirichlet boundary condition.}. On the other hand, with a temperature between $T_c(\tilde{\rho}\rightarrow{\cal R})$ and $T_c(\tilde{\rho}= 0)$, the system becomes spatially inhomogeneous with a confined region at the core and a deconfined state in the outer layer. Our result agrees with the finding in Ref.~\cite{Chernodub:2020qah}, while lattice simulations for $SU(3)$~\cite{Braguta:2023iyx} and caloron model study for $SU(2)$~\cite{Jiang:2024zsw} give an opposite tendency, where $T_c$ gets increased for increasing ${\tilde \rho}$. Despite the qualitative difference, we find that similar to Ref.~\cite{Braguta:2023iyx}, for relatively small ${\tilde \rho}$, a quadratic fit
\beq
\label{fit1}
T_c(\tilde{\rho})=T_c(\tilde{\rho}=0)-{\rm c}^\prime \tilde{\rho}^2\, ,\quad{\rm with}\quad {\tilde \rho}/{\cal R}\le 0.5\,, 
\eeq
can well reproduce the ${\tilde \rho}$-dependence of $T_c$ for any fixed $\Omega$. In our case, the $\Omega$-dependent number ${\rm c}^\prime$ is positive, while a negative number is required to fit the lattice data. 

The above observation can be also attributed to the centrifugal effects as already mentioned above. We remark that as a consequence of the finite-volume effect, even at $\Omega=0$, the deconfining temperature $T_c$ does not revert to $T_d$; instead, $T_c$ exhibits a ${\tilde \rho}$-dependence and is always lower than $T_d$. This is in agreement with the lattice simulations~\cite{Lucini:2002ku} of a non-rotating system, performed on a hypercubic lattice with periodic boundary conditions. However, for a sufficiently large $\mathcal{R}$, we can expect a negligible finite-volume effect so that the critical temperature $T_c$ becomes very close to $T_d$ at radial positions near the rotation axis. On the other hand, instead of fixing the radial size ${\cal R}$ as we did in this work, one can also take ${\cal R} \Omega$ to be fixed~\cite{Chernodub:2020qah}. In the latter case, taking $\Omega\rightarrow 0$, $T_c$ becomes ${\tilde \rho}$-independent and identical to $T_d$ as $\mathcal{R}\to \infty$. This has already been proved in Sec.~\ref{MMO-in-real-rotation}. Furthermore, we also find that with a $\Omega$-dependent radial size, the corresponding results for $\Omega \neq 0$ qualitatively agree with those presented in Figs.~\ref{TTdrTdvalueRe} and \ref{TTdwTdvalueRe}. From a quantitative point of view, notable changes arise at very small and large angular velocities, where the volume size becomes extreme relative to our choice.

\begin{figure}[htbp]
    \centering    \includegraphics[width=0.45\textwidth]{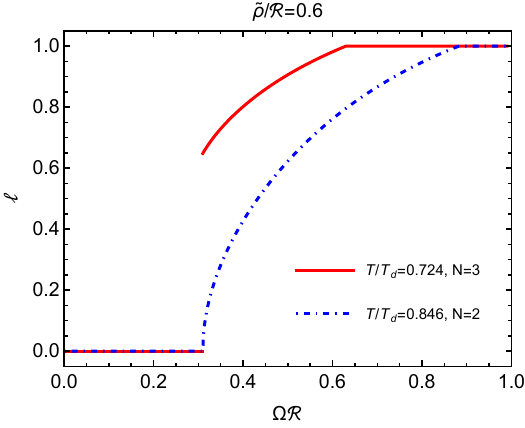} 
\hfill 
\includegraphics[width=0.45\textwidth]{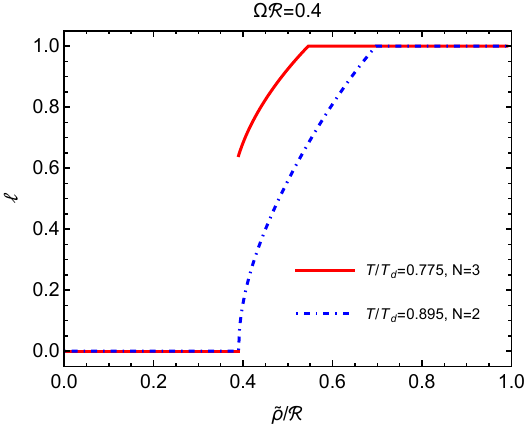} 
        \caption{The Polyakov loop $\ell$ as a function of $\Omega\mathcal{R}$ at fixed $\tilde{\rho}/\mathcal{R}$ (left panel) and as a function of $\tilde{\rho}/\mathcal{R}$ at fixed $\Omega\mathcal{R}$ (right panel).}
        \label{LwTd04}
\end{figure}

Fig.~\ref{LwTd04} shows the Polyakov loop $\ell$ as a function of $\Omega\mathcal{R}$ at fixed radial position, and as a function of $\tilde{\rho}/\mathcal{R}$ at fixed angular velocity. In these plots, the specific values of the temperature are chosen so that the critical angular velocity (left panel) and the critical radius\footnote{It severs as a phase boundary of the mixed phase.} (right panel) become identical for both $SU(2)$ and $SU(3)$. According to our results, one can expect that with fixed $\Omega$ and $\tilde{\rho}$, a temperature closer to the corresponding $T_d$ in the infinite-volume limit is required to drive the deconfining phase transition for $SU(2)$, compared to $SU(3)$. In addition, beyond the critical radius and the angular velocity, we can see a rather rapid increase of $\ell$  with the increased $\Omega$ and $\tilde{\rho}$, which is related to a dramatic decrease of the background field toward zero. However, a decreased $\ell$ is observed when analytically continuing the lattice result from imaginary to real angular velocity~\cite{Braguta:2020biu, Braguta:2021jgn,Braguta:2021ucr,Braguta:2023iyx}. It should be noted that the behavior of $\ell$ as shown in Fig.~\ref{LwTd04} is actually consistent with the Tolman-Ehrenfest law and supports the fact that rotation accelerates the entry into the deconfinement. Our results also demonstrate that neither rotation nor finite-volume effects alter the nature of the phase transition, it is first order for $SU(3)$ and second order for $SU(2)$.  

\subsubsection{$M$ depends on $\Omega$}\label{mgw}

To avoid extra assumptions in the matrix model, the above results are based on the use of a constant effective mass $M$. However, even for a non-rotating system in the infinite-volume limit, one can consider some possible $T$-dependence of the effective mass as did in Ref.~\cite{Guo:2014zra}. The situation becomes even more complicated when discussing a rotating plasma in a cylinder because, in principle, $M$ could be an involved function of several variables including $T$, ${\tilde \rho}$ and $\Omega$. We assume $M(\Omega)=g(\Omega)T$, so that the effective mass acquires its $\Omega$-dependence through the coupling constant $g(\Omega)$. Following Refs.~\cite{Jiang:2021izj, Jiang:2023zzu}, we take $g(\Omega)=(1+0.1 \Omega/\Lambda)g_0$ where $g_0$ is the coupling constant at vanishing $\Omega$ and $\Lambda = 0.28$ {\rm GeV}. Our assumption inevitably introduces a non-trivial $T$-dependence in the effective mass and we will simply ignore it since our focus will be on the $\Omega$-induced correction to the effective mass. Therefore, $M(\Omega=0)$ is set equal to the previously used constant $M$. As a result, a specific form $M(\Omega)=(1+0.1 \Omega/\Lambda)M$ will be adopted in the following.

\begin{figure}[htbp]
    \centering
\includegraphics[width=0.45\textwidth]{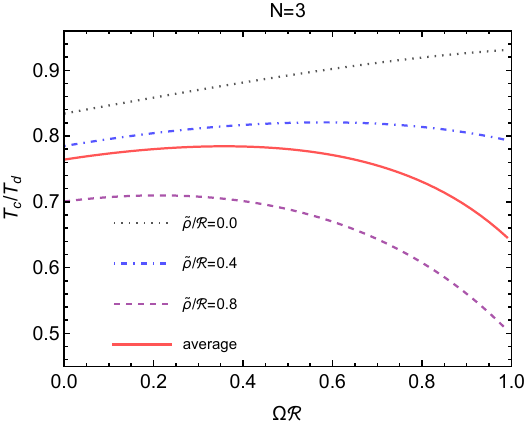} 
\hfill 
\includegraphics[width=0.45\textwidth]{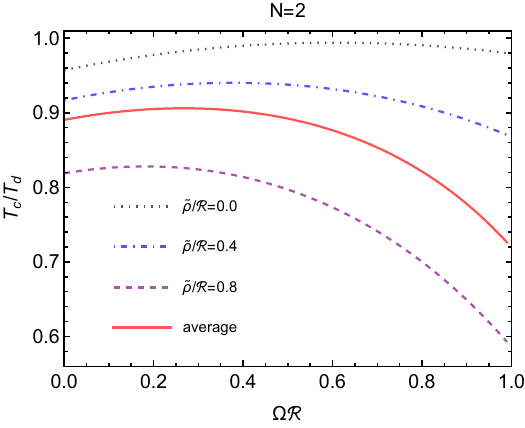} 
\caption{The temperature ratio $T_c/T_d$ as a function of $\Omega \mathcal{R}$ with $\Omega$-dependent effective mass.}
    \label{TTdrTdvaluewRe}
\end{figure}

Given the $\Omega$-dependent effective mass, we show $T_c/T_d$ as a function of $\Omega \mathcal{R}$ in Fig.~\ref{TTdrTdvaluewRe}. Compared with Fig.~\ref{TTdrTdvalueRe}, the most notable difference is that $T_c$ exhibits a non-monotonic dependence on $\Omega$. When $\Omega$ is small, $T_c$ increases with increasing $\Omega$ which is qualitatively in agreement with lattice simulations~\cite{Braguta:2020biu, Braguta:2021jgn,Braguta:2021ucr,Braguta:2023iyx}. However, a quadratic fit as given in Eq.~(\ref{fit2}) doesn't work well to reproduce the results in this figure. While in the large $\Omega$ region, it shows a decrease with $\Omega$. We point out that with the caloron model~\cite{Jiang:2023zzu}, the same non-monotonic behavior of $T_c$ appears when the same $\Omega$-dependent coupling constant is used. However, only a special case of $\Tilde{\rho}=0$ for $SU(2)$ was considered in this model. In the bag model~\cite{Mameda:2023sst}, the $\Omega$-dependence of $T_c$ is found to differ depending on whether or not the bag constant is taken to be $\Omega$-dependent. This agrees with our different results in the small $\Omega$ region as shown in Fig.~\ref{TTdrTdvalueRe} and Fig.~\ref{TTdrTdvaluewRe}. Notice that the non-monotonic behavior is not observed in Ref.~\cite{Mameda:2023sst} because the corresponding results are only available for small angular velocities.  

In our model, such a non-monotonic behavior of $T_c$ can be understood as a competition between two opposite effects. As $\Omega$ increases, on the one hand, the confining effect gets enhanced due to the increased effective mass $M(\Omega)$ which is associated with the non-perturbative dynamics of the gluon plasma, the system tends to stay in the confined phase, indicating a higher $T_c$. On the other hand, with further increased $\Omega$, the rotation-induced centrifugal effect that facilitates deconfinement becomes dominant, consequently, we find a decreased $T_c$. It is worth noting a notable exception, for $SU(3)$, the deconfining temperature along the rotation axis increases monotonically with $\Omega$. This is actually consistent with the observation in Fig.~\ref{TTdrTdvalueRe} where $T_c({\tilde \rho}=0)$ is nearly independent of $\Omega$, showing a negligible centrifugal effect. As a result, only the enhanced confining effect plays a role, leading to an increased $T_c$.

\begin{figure}[htbp]
    \centering
\includegraphics[width=0.45\textwidth]{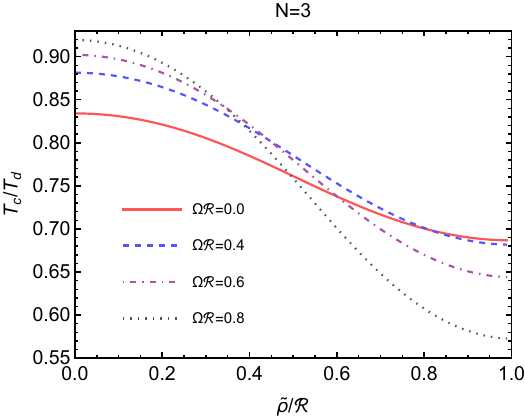} 
\hfill 
\includegraphics[width=0.45\textwidth]{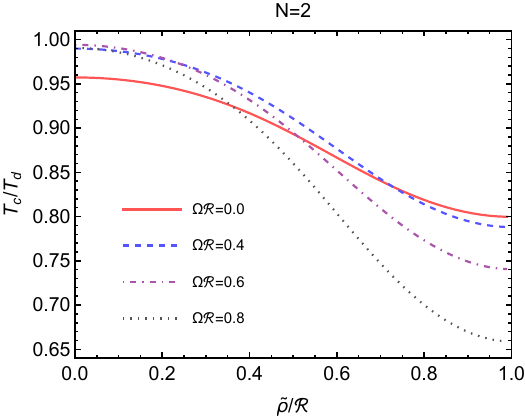} 
\caption{The temperature ratio $T_c/T_d$ as a function of $\tilde{\rho}/\mathcal{R}$ with $\Omega$-dependent effective mass.}
    \label{TTdwTdvaluewRe}
\end{figure}

We also present $T_c/T_d$ as a function of $\tilde{\rho}/\mathcal{R}$ in Fig.~\ref{TTdwTdvaluewRe}. With both constant $M$ and the $\Omega$-dependent one, the $\rho$-dependence in the matrix model only lies in the Bessel function of the first kind. Therefore, we can naturally expect a similar behavior when compared with the results in Fig.~\ref{TTdwTdvalueRe}. This is confirmed in Fig.~\ref{TTdwTdvaluewRe} where $T_c$ decreases with increasing $\Tilde{\rho}$. In addition, we find that a quadratic fit similar to Eq.~(\ref{fit1}) also works well to reproduce the results in this figure for small ${\tilde \rho}/{\cal R}$.

\begin{figure}[htbp]
    \centering    \includegraphics[width=0.45\textwidth]{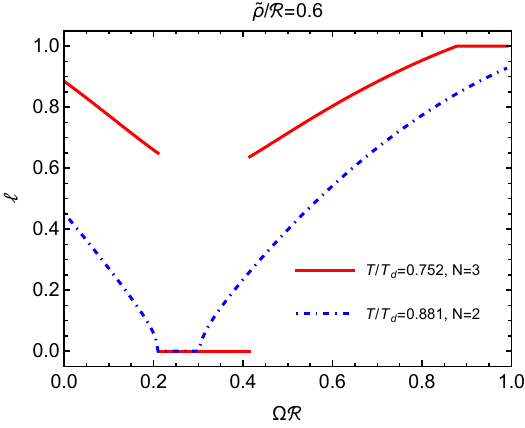} 
    \includegraphics[width=0.45\textwidth]{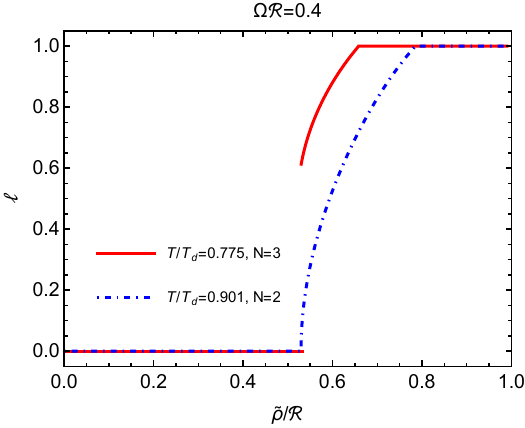} 
        \caption{The Polyakov loop $\ell$ as a function of $\Omega\mathcal{R}$ at fixed $\tilde{\rho}/\mathcal{R}$ (left panel) and as a function of $\tilde{\rho}/\mathcal{R}$ at fixed $\Omega \mathcal{R}$ (right panel). The results are obtained with $\Omega$-dependent effective mass.}
        \label{LwTd04w}
\end{figure}

Finally, we briefly comment on the behavior of the Polyakov loop $\ell$ when using the $\Omega$-dependent effective mass which can be found in Fig.~\ref{LwTd04w}. Given $\Omega$ and $T$, we find that $\ell$ increases with $\tilde{\rho}$, provided that $\tilde{\rho}$ is larger than the critical radius that separates the confined and deconfined phases. So it has no qualitative difference as compared with Fig.~\ref{LwTd04}. On the other hand, with fixed $\tilde{\rho}$ and $T$, $\ell$ becomes a non-monotonic function of $\Omega$ and there exists two critical angular velocities $\Omega_c$. Starting from a deconfined phase and increasing $\Omega$ from zero, the system undergoes a transition into a confined phase when the angular velocity reaches the smaller critical value $\Omega_c$. It then remains confined until $\Omega$ increases to the larger $\Omega_c$, where a confinement-to-deconfinement phase transition occurs. Such a behavior is a direct consequence of the non-monotonic $\Omega$-dependence of $T_c$ shown in Fig.~\ref{TTdrTdvaluewRe}. Therefore, a similar $\Omega$-dependence of $\ell$ can be expected based on the caloron model~\cite{Jiang:2023zzu}. On the other hand, lattice simulations seem not to support this kind of behavior because $T_c$ monotonically increases with real-valued angular velocity~\cite{Braguta:2020biu, Braguta:2021jgn,Braguta:2021ucr,Braguta:2023iyx}. However, to check whether or not there is a turning point at which $T_c$ starts to fall with increasing $\Omega$, according to Fig.~\ref{TTdrTdvaluewRe}, one needs to go to large-$\Omega$ region where, however, analytical continuation from imaginary- to real-valued angular velocity becomes not reliable.
 
\section{Summary and Outlooks} \label{summary}

In this work, we extended the matrix model for deconfinement to a rotating system, confined in a cylinder with a finite radius ${\cal R}$. In the frame of finite temperature field theory, rotation effect was introduced by computing the partition function in cylindrical coordinates that corotate with the system at a constant angular velocity $\Omega$, satisfying ${\cal R} \Omega < 1$. The partition function was obtained by solving the eigenvalue problems of scalar and vector fluctuation operators in which both the ghost and gluon field acquired an effective mass $M$. By retaining the first two terms in the high temperature expansion of the effective potential, we constructed the matrix model for a rotating gluon plasma which exhibited a radial inhomogeneity.

The matrix model becomes homogeneous when taking $\mathcal{R}\to \infty$ and $\Omega=0$, and thus reproduces the original matrix model for a non-rotating plasma in an infinite volume. Despite its complex form involving infinite sums of Bessel functions, the matrix model under real rotation shares a phase transition mechanism similar to that of the original matrix model, in which the transition arises from the competition between the perturbative and non-perturbative contributions that constitute the model. We have analytically proved that the minimum of the perturbative term is always located at ${\sf q}=0$, indicating a completely deconfined gluon plasma; while the non-perturbative contribution reaches its minimum at ${\sf q}=1/3$ for $SU(3)$ and at ${\sf q}=1/4$ for $SU(2)$, therefore, favors confinement. These features are exactly the same as those found in the original matrix model and real rotation only changes the stabilities of these vacua.

Using the matrix model under real rotation, we numerically investigated the behaviors of the deconfining temperature $T_c$ and the Polyakov loop $\ell$ for both $SU(3)$ and $SU(2)$. In general, the deconfining temperature is reduced in a rotating plasma as compared to that for a non-rotating system in an infinite volume, indicating that real rotation accelerates the phase transition to deconfinement. For the case of a constant effective mass $M$, we found that $T_c$ decreases as $\Omega$ increases, and this trend can be well fitted by a quadratic function of $\Omega$ in the small $\Omega$ regime. Furthermore, our result also showed that $T_c$ decreases with increasing $\tilde{\rho}$ and exhibits a quadratic dependence on $\tilde{\rho}$ near the rotation axis. As a result, the rotating system becomes spatially inhomogeneous with a confined region at the core and a deconfined state in the outer layer, which is consistent with the Tolman-Ehrenfest law. The decrease of $T_c$ with both $\Omega$ and ${\tilde \rho}$, being particularly marked at large $\Omega$ and ${\tilde \rho}$, can be attributed to the centrifugal effect generated by the real rotation. On the other hand, by setting $\Omega=0$, a finite-volume effect on the deconfining temperature was observed. Due to the small radial size ${\cal R}$ used in our evaluations, such an effect becomes significant especially near the boundary and leads to a reduced $T_c$ compared to that in the infinite-volume limit. We also studied the $\Omega$- and $\Tilde{\rho}$-dependence of the Polyakov loop $\ell$ at fixed temperatures. Our results revealed the existence of the critical values of angular velocity and radial position below which $\ell$ vanishes. Beyond these thresholds, $\ell$ rises rapidly with $\tilde{\rho}$ and $\Omega$.  
In addition, the behavior of $\ell$ clearly showed that the phase transition is first order for $SU(3)$ and second order for $SU(2)$, which is not
altered by the rotation and finite-volume effect.

Taking into account a possible $\Omega$-dependence of the effective mass $M$, the most notable result is that $T_c$ shows a non-monotonic dependence on $\Omega$. As compared to the case with constant $M$, a qualitative change was found in the small-$\Omega$ region where $T_c$ gets increased with increasing $\Omega$. Accordingly, a decreased $\ell$ was observed in this region.
Such a behavior can be understood as an accentuated confining effect due to the increased effective mass which is dominant over the centrifugal effect in the small-$\Omega$ region. On the other hand, for large angular velocities, the centrifugal effect plays a dominant role, which in turn leads to a decrease of $T_c$.

We have also conducted comprehensive comparisons between our results and those from various lattice simulations and phenomenological models. However, controversy remains over how the properties of the deconfining phase transition is modified by real rotation. Not only do contradictions exist between lattice simulations and model studies, but there are also inconsistencies within each of these approaches. Therefore, further work is still needed to reach a definite conclusion. 
Despite the above issues, incorporating the fermionic contributions in the matrix model to examine the phase structure of a rotating QGP is certainly an interesting avenue for future work.

\section*{Acknowledgements}
We thank Yin Jiang for useful discussions. The research of YG is supported by the National Natural Science Foundation of China (NSFC) with Grant No. 12465022; of QD by the NSFC with Grant No. 12305135, by Guangxi Natural Science Foundation with Grant No. 2023GXNSFBA026027 
and by Guangxi Young Elite Scientist Sponsorship with Program No. GXYESS2026074; of MH by the NSFC with Grant Nos. 12235016 and 12221005, and of EW by the NSFC with Grant No. 12635009. 

\bibliography{rotation_deconfinment}

\end{document}